%% file: FrischSchwadron-101013-astroph.tex
\documentclass[12pt]{article}
\usepackage{wrapfig}
\usepackage{natbib}
\usepackage{epsfig}

\newcommand\aj{{AJ}}%
\newcommand\araa{{ARA\&A}}%
\newcommand\apj{{ApJ}}%
\newcommand\apjl{{ApJ}}%
\newcommand\apjs{{ApJS}}%
\newcommand\aap{{A\&A}}%
\newcommand\mnras{{MNRAS}}%
\newcommand\ssr{{Space~Sci.~Rev.}}%

\begin{document}
\def\cc{cm$^{-3}$}
\def\cmtwo{cm$^{-2}$}
\def\kms{km s$^{-1}$}

\begin{center}
\begin{Large}
{\bf Large-scale Interstellar Structure and the Heliosphere} \\
\end{Large}
\vspace*{0.02in}
{P. C. Frisch, and N. A. Schwadron}\footnote{Affiliations:  
PCF--University of Chicago, Chicago, IL  60637;
NAS--University of New Hampshire, Durham, NH  03824}
\end{center}

\begin{abstract}

The properties of interstellar clouds near the Sun are ordered by the
Loop I superbubble and by the interstellar radiation field.
Comparisons of the kinematics and magnetic field of the interstellar
gas flowing past the Sun, including the Local Interstellar Cloud
(LIC), indicate a geometric relation between Loop I as defined by
radio synchrotron emission, and the interstellar magnetic field that
polarizes nearby starlight.  Depletion of Fe and Mg onto dust grains
in the LIC shows a surprising relation to the far ultraviolet
interstellar radiation field that is best explained by a scenario for
the LIC to be extended, possibly filamentary, porous material drifting
through space with the Loop I superbubble.  The interstellar velocity
and magnetic field measured by the Interstellar Boundary Explorer
(IBEX) help anchor our understanding of the physical properties of the
nearby interstellar medium.

\end{abstract}

\section{Introduction}

The first maps of the distribution of interstellar dust, obtained from
starlight reddened in the interstellar medium (ISM), showed that the
Sun is located in a void with a radius of $\sim 75$ pc
\citep{Fitzgerald:1968}.  This void is now known as the ``Local
Bubble''.  The dawn of the space age enabled the discovery that
hydrogen and helium flow through the heliosphere with the velocity of
nearby interstellar gas
\citep[][]{BertauxBlamont:1971,ThomasKrassa:1971,WellerMeier:1974,AdamsFrisch:1977},
and that low density partially ionized gas is seen toward the closest
stars \citep{RogersonYork:1973}.  These earliest data showed that the
Sun is embedded in a low density partially ionized interstellar cloud,
the Local Interstellar Cloud (LIC), with low volume densities ($<<1$
\cc).  Recent studies indicate that hydrogen is 23\% ionized, and
helium is $\sim 39$\% ionized in the LIC, and that refractory elements
have relatively high gas-phase abundances
\citep[][FRS11]{Frisch:2011araa}.  In the global ISM, clouds such as
the LIC were first termed the ``intercloud medium'' because they are
warm and optically thin and favor higher velocities, $> 20$ \kms
\citep{Welty23:1999}, through the local standard of rest velocity
frame (LSR); LIC-like clouds are not at risk of gravitational
collapse, in contrast to the local cold, opaque, dense low velocity
clouds that border the Local Bubble.  Recent discoveries of tiny dense
compact clouds through observations of the H$^\circ$ 21-cm hyperfine
transition \citep[e.g.][]{SaulGALFAPeek:2012sss}, even within $\sim
20$ pc of the Sun \citep{MeyerPeek:2012leo}, and the relation between
Loop I and the interstellar magnetic field (ISMF) and gas around the
heliosphere \citep{Frisch:1981,Frisch:2012ismf2}, signals that the
local environment is a complex interstellar region.  The location of
the Sun in the Local Bubble void allows the interstellar radiation
field (ISRF) to affect the LIC.

\section{Cloud kinematics, Loop I and the local interstellar magnetic field }

Local interstellar clouds are identified by absorption lines formed in
the spectra of nearby stars by the foreground interstellar
gas. High-spectral resolution spectra are interpreted using the
assumption that the velocities of the atoms forming the absorption
\input{frischfig1-astroph} lines have Maxwellian velocity
distributions in the cloud, with an additional ``turbulent''
broadening to explain mass-independent non-thermal motions.
Ultimately, the detailed kinematical structure of the local
interstellar medium that is derived from absorption line data depends
on the spectral resolution of the measuring instrument, which is
generally $>3$ \kms\ for ultraviolet (UV) data and $> 1$ \kms\ for
ground-based data.  For optical Na$^\circ$, Ca$^+$, and K$^\circ$
lines, the frequency of adjacent components with small separations in
velocity in a given sightline increases significantly as separations
become smaller, but decreases near the limit of instrumental spectral
resolutions \citep{WeltyK:2001}.  This indicates that the UV data that
provide most of our information on local clouds under-samples the
velocity structure of the local ISM.

The distribution of cloud velocities obtained from interstellar
absorption components in the spectra of nearby stars indicate that the
Sun is located in a decelerating flow of interstellar gas
\citep{FGW:2002}. Converting the observed heliocentric velocities into
the LSR gives a bulk flow velocity with an LSR flow speed of --17
\kms\ from the ``true'' LSR upwind direction that is toward the
direction $\ell, b \sim 335^\circ,-5^\circ$ (galactic
coordinates, FRS11).  The kinematic structure of the
local ISM has been parsed by RL08 into fifteen nearby clouds (earlier
studies had derived fewer clouds).  Figure 1 shows the LSR velocities
of the RL08 clouds that we have derived using the solar apex motion in
\citet[][see Table 1]{SchonrichBinneyDehnen:2010}.

The deceleration of the bulk motion of nearby ISM produces regions of
cloud collisions \citep[e.g.][RL08]{RLIV}.  The closest of those
collisions is between the LIC and Blue Cloud (BC), which have LSR
velocity vectors that are separated by an angle of $47^\circ$ (Table
1), and a relative speed between the two clouds of $|V|\sim 10.5$
\kms.  For a LIC density of 0.27 \cc\ \citep[model 26
in][SF08]{SlavinFrisch:2008}, and reduced mass per particle of $2.34
\times 10^{-24}$ gr, the ram pressure of the LIC on the BC is
$P_\mathrm{ram} = 1.28 \times 10^{−12}$ dynes \cmtwo. This
is a factor of five larger than the LIC thermal pressure of $2.32
\times 10^{−13}$ dynes \cmtwo, using the LIC temperature
6300 K from \citet{McComas:2012bow}, indicating a supersonic interface
between the LIC and BC.  Nearby sources of intraday
scintillation sources have been identified close to the direction of
the LIC-BC collision \citep{Linsky_etal_2008}.

Loop I is an evolved superbubble originally identified by an extended
arc of synchrotron emission \citep[e.g.][]{Haslametal:1971}.  Models
of the Loop I shell, based on the hydrodynamic expansion of a
superbubble into low density ISM, explain the large size and give an
age of 4 Myrs, corresponding to the approximate age of the latest
epoch of star-formation in the Scorpius-Centaurus Association that
formed the Upper Scorpius subgroup \citep{Frisch:1995rev}.
Measurements of starlight polarized by interstellar dust show that the
east and west sides of Loop I differ in the distance of polarizing grains \citep{Santosetal:2010}.
\citet{Wolleben:2007} has separated Loop I into two spherical shells;
the Sun is located in the rim of the S1 component (gray arc in Figure
1) that is centered $78 \pm 10$ pc away at galactic coordinates of
$\ell,b=346^\circ \pm 5^\circ,3^\circ \pm 5^\circ$.

The S1 shell has a notable relation to the properties of the very
local ISM: (1) the bulk motion of nearby clouds in the LSR is directed
away from a direction within $14^\circ \pm 8^\circ $ of the S1 shell
center (thick gray arrow in Figure 1; the thin arrows originate at
cloud centers and the distance of the nearest cloud star).  (2) the
local ($<40$ pc) ISMF direction determined from weighted fits of
starlight polarized by magnetically aligned dust, $\ell, b \sim
47^\circ \pm 20^\circ ,25^\circ \pm 20^\circ$, makes an angle of $\sim
76^\circ \pm 21^\circ$ with the bulk flow LSR velocity
\citep{Frisch:2012ismf2}.  Both properties are expected for ISM
associated with an expanding superbubble shell that has the ISMF
perpendicular to a normal to the shell surface (e. g. aligned with
that surface).

In addition to being the home of the heliosphere (for now at least),
the LIC is special since the heliosphere itself probes the ISM.  The
Interstellar Boundary Explorer (IBEX) measured a Ribbon of energetic
neutral atoms, with energies 0.2--6 keV, which appears to form in
directions perpendicular to the ISMF as it drapes over the heliosphere
\citep{McComas:2009,Schwadron:2009}.  The Ribbon center at 2.73 keV is
a proxy for the ISMF direction at the heliosphere, $\ell = 30.5^\circ
\pm 2.6^\circ$, $b=57.1^\circ \pm 1^\circ$ \citep[with the
uncertainties from the variation of the ribbon center with
energy,][]{Funsten:2013}.  The angle of $33^\circ \pm 20^\circ$
between the very local ISMF obtained from the IBEX Ribbon and the ISMF
determined over 8--40 parsec scales from polarized starlight suggests
the two directions may be measuring different regions of the same
ISMF.  IBEX has also measured the velocity of interstellar He$^\circ$
in the inner heliosphere, finding a value that closely agrees with the
LIC velocity obtained from spectroscopy \citep[][Table
1]{McComas:2012bow}.  The IBEX measurements of the ISM kinematics and
ISMF at the heliosphere anchor our understanding of the LIC.

Photoionization models of the LIC give the LIC physical properties
n(H$^\circ) \sim 0.19$ \cc\ and n(H$^+ ) \sim 0.07$ \cc\ (SF08); for
equality of thermal and magnetic pressures, the magnetic field
strength in the LIC would be $\sim 3$ $\mu$G.  A similar estimate for
the field strength arises from pressure balance based on the
line-of-sight integrated pressure observed in energetic neutral atoms
by IBEX \citep{Schwadronetal:2011sep}. In global low electron density
regions, Faraday Rotation suggests that density fluctuations and
magnetic field strength are uncorrelated \citep{WuKimRyu:2009}, in
agreement with the lack of evidence for flux freezing for clouds with
densities $< 10^3$ \cc\ \citep{Crutcher:2007flxfrz}.

\section{Abundances, depletions, and the interstellar radiation field }

The gas-phase abundances of refractory elements such as Fe and Mg are
a key diagnostic of the history of a cloud because grain erosion by
sputtering and collisions with other grains in interstellar shocks
return the refractory elements to the gas phase.  It has long been
known that the abundances of refractory elements in the gas of clouds
with higher velocities are larger than those in cold slow dense clouds
\cite[e.g.][]{Welty23:1999}.  The LIC provides the ISM sample that is
most likely to be uniform for the purpose of understanding this
effect.  In the local ISM, the variations of the ratio Fe$^+$/Mg$^+$
in the gas within and between clouds suggest that interstellar shocks
are presently active in destroying local dust grains (RL08,FRS11).
This dust should consist of grains that were swept up and processed by
the superbubbles that form the Loop I shell as it propagated through
space \citep{Frisch:1995rev}, but an admixture of evaporated dense
cloud material is needed to explain the large micron-sized grains
observed by Ulysses (FRS11). 

Except for the LIC where photoionization corrections provide the
amount of H$^+$ present, most local ISM abundances are found based
only on H$^\circ$ column densities.  This comparison yields erroneous
abundances for the common species such as Fe$^+$ and Mg$^+$ that trace
both the neutral and ionized components of warm partially ionized
material (WPIM, SF08).  Elemental gas-phase abundances are therefore
overestimated in WPIM if ionization corrections are ignored, or
equivalently the number of atoms in the grains are
underestimated. Elemental ``depletions'' are given by $\delta_X =
\mathrm{log}~[N(X)/N(H)|_\mathrm{is}/N(X)/N(H)|_\mathrm{sun} ]$, where
the first and second terms give the interstellar and solar abundances
of X=Fe$^+$ or X=Mg$^+$ with respect to the column density N(H)
\citep[e.g. see][]{Welty23:1999}.  If the neglect of H$^+$ biases
abundances in the local ISM, these biases should be more prominent in
regions of high than low radiation fluxes.

Figure 2, top panel, shows the Fe and Mg depletions for eight stars,
d$<30$ pc, with LIC components (from RL08), plotted against the
far-ultraviolet (FUV) radiation flux at the position of each LIC star.
The radiation field at 975\AA\ is used for this comparison
\citep[][Figure 12 of Frisch et al. 2012]{OpalWeller:1984}.
\nocite{Frisch:2012ismf2} The radiation field at 912\AA, the
wavelength corresponding to the first ionization potential of
H$^\circ$, would be preferable to use but those data are not
available; measurements of the brightest stars at 975\AA\ are used as
a proxy.  The proxy stars include the brightest EUV sources $\epsilon$
CMa and $\beta$ CMa that dominate the LIC ionization at the Sun
(SF08), and consist mainly of luminous B1 through O stars located
where $\ell > 180^\circ$ and EUV interstellar opacities are very low
due to the Local Bubble.  HD 122451 is the brightest star at 975\AA\
and it is located in the direction of the Loop I center.  The 975\AA\
radiation field at the Earth is useful as a proxy for the H-ionizing
radiation (912\AA) at nearby stars if the LIC is porous so that cloud
opacity does not decouple the 975\AA\ and 912\AA\ fluxes, and if the
LIC is extended so that the inverse square dependence of the radiation
\input{frischfig2-astroph} field is significant over the length of the
cloud.  Local clouds typically fill less than 10\%--30\% of space
(FRS11), so these assumptions are plausible.  The radiation fluxes are
calculated with the assumption of a transparent ISM at 975\AA\ between
the LIC star and each 975\AA\ radiation source.  Figure 2, top, shows
a clear trend for the amount of Fe and Mg in dust grains in the LIC to
decrease as the FUV radiation flux at the star increases, providing
that the amount of ionized hydrogen present is insignificant.

The plotted depletions of Fe and Mg in the LIC show that the ISRF
($F_{975}$) is proportional to $\delta_\mathrm{Fe}$ and
$\delta_\mathrm{Mg}$, so that abundances in the gas phase increase
with the radiation flux.  The black and blue dashed lines show linear
fits to the depletion versus flux trend for Fe$^+$ and Mg$^+$,
respectively, giving iron depletions $\delta_\mathrm{Fe} = -3.0 + 0.29
*F_{975}$ with a reduced $\chi^2$ of 0.8, and magnesium depletions
$\delta_\mathrm{Mg} = -4.0 + 0.46*F_{975}$ with a reduced $\chi^2$ of
3.8.  The linear fit to the Fe depletion-flux variation is better than
found for Mg.  Mg equilibrium in warm partially ionized gas is more
complicated, since dielectric recombination of Mg$^+$ with an electron
is significant, and Mg$^{++}$ forms in regions of high radiative flux
(e.g.  15\% of Mg is doubly ionized at the solar location, SF08).  The
dependence of the depletions on the ISRF could result from an increase
in gas ionization with increasing radiation flux, which when ignored
would yield values of $\delta_\mathrm{Fe}$ and $\delta_\mathrm{Mg}$
that are too large.  An alternative, that H$_2$ is present and
increases with depth in the cloud, is unlikely because H$_2$ is easily
photodissociated by UV radiation in environments such as the LIC.
Another possible explanation for the depletion-radiation relation is
that the collision between the LIC and BC, discussed above, creates
shocks that preferentially destroy the grains in the region of the
collision. Since the BC is located in the direction of the bright EUV
sources $\epsilon$ CMa and $\beta$ CMa, this would resemble a
dependence on radiation fluxes.  The option that cloud opacity
accounts for the depletion-radiation relation does not work if the LIC
is a compact cloud restricted to being within a parsec of the
heliosphere, since the depletion variation is plotted with respect to
the distance between the LIC star and the radiation source.

A more interesting possibility that is consistent with Figure 2 is
that the LIC gas consists of extended porous material, drifting
through space at the same velocity, as part of the structure of the
Loop I superbubble shell. For this configuration, the column of ISM
between the star and radiation sources would see the largest gradient
in flux.  This scenario is consistent with the solar location at the
upwind edge of the LIC (FRS11), and explains why the
depletion-radiation effect is found at larger distances than is
expected if the LIC were compact.

The middle panel in Figure 2 shows an example of the LIC fractional
ionization, H$^+$/(H$^\circ$+H$^+$), needed to explain the depletion
variation with radiation flux.  Ionization is calculated assuming an
arbitrary constant value for the base-line depletion.  The best fits
through the ionization variation indicates variations larger than 50\%
in the LIC.

The bottom panel in Figure 2 plots the LIC temperature (from RL08)
against radiation flux for these sightlines.  At the Sun, $\sim 66$\%
of the cloud heating is caused by absorption of photons that can
ionize H$^\circ$ and He$^\circ$ (SF08).  Stars in the third and fourth
galactic quadrants ($\ell > 180^\circ$, open circles) are clustered in
high flux regions and tend to have higher temperatures than the
remaining stars. The remaining stars appear to warm as the ISRF
decreases, possibly due to decreased cooling by the collisionally
excited fine-structure lines of heavy elements.

\section{Summary}

The Loop I evolved superbubble orders the kinematics of nearby
interstellar clouds and the local magnetic field.  The bulk flow LSR
velocity of the local cloud system is perpendicular to the ISMF,
making an angle of $76^\circ \pm 21^\circ $ with the field direction
found from starlight polarized by the ISM.  The LSR velocity is also
aligned with the geometric center of Loop I, to within $14^\circ \pm
8^\circ $, as defined by the S1 shell seen in synchrotron emission.  A
surprising relation between the depletion of Fe and Mg onto dust
grains in the LIC, and the far ultraviolet interstellar radiation
field at 975\AA, is best explained by a scenario for the LIC to be
extended, possibly filamentary, porous material drifting through space
with the near side of Loop I.  The eastern arc of Loop I in galactic
quadrant I ($\ell < 90^\circ$) appears to extend to the solar
location, while the properties of the ISM in the opposite direction
(quadrant III, $180^\circ < \ell < 270^\circ$) are dominated by the
FUV interstellar radiation field.

\vspace*{-0.01in}

{\it Acknowledgements} This work has been supported by the
Interstellar Boundary Explorer mission as a part of NASA’s
Explorer Program.  To be published
in the proceedings of the 12th Annual International
Astrophysics Conference ``Outstanding problems in heliophysics: from
coronal heating to the edge of the heliosphere'', Editor Qiang Hu.

\input{frischtable1-100813-astroph}
\vspace*{-0.15in}
%\bibliographystyle{asp2010}
%\bibliography{ges2}

\end{document}

%% file: frischfig1-astroph.tex
\begin{wrapfigure}{l}{8.0truecm}
  \vspace{-0.75cm}
  \hspace{-1.1cm}
  \centering
  \includegraphics[width=8.6cm]{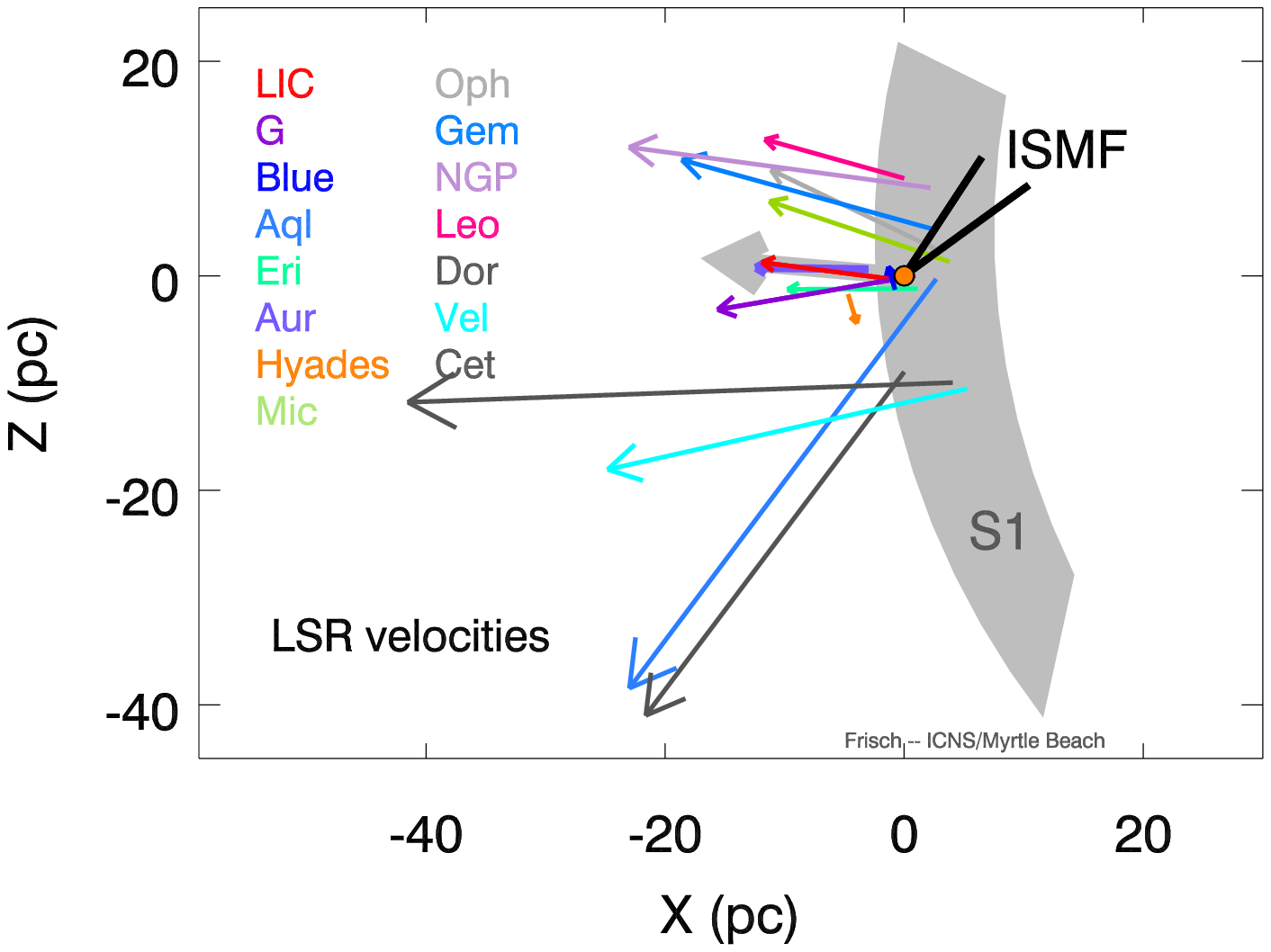}
\caption{Loop I and local clouds.  Arrows show LSR velocities of local interstellar clouds;
the thick gray arrow shows the bulk LSR motion of the local cloud system.
The X/Z axes are aligned with the galactic center/north pole.
The upper and lower black lines give the directions of the ISMF obtained from the
IBEX ribbon and starlight polarizations, respectively,
projected onto the X-Z plane. The orange dot indicates the Sun 
(details in text).  } \label{fig:s1}
\end{wrapfigure}

%% file: frischfig2-astroph.tex
\begin{wrapfigure}{l}{10.0truecm}
%\begin{figure}{h!t!}
\centering
\hspace*{-0.6cm}
\includegraphics[width=10.0truecm]{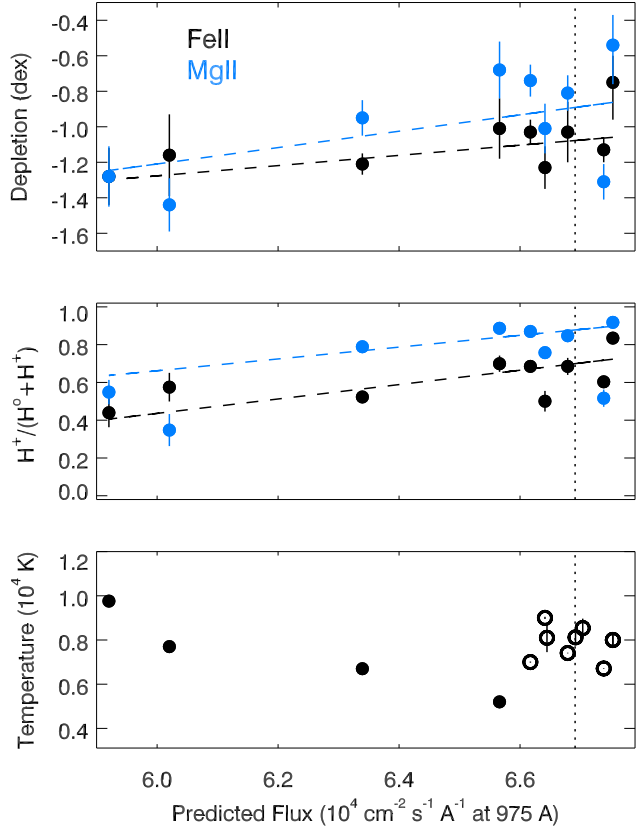}
\caption{Top:  Depletions $\delta_\mathrm{Fe}$ (black) and $\delta_\mathrm{Mg}$
(blue) in the LIC for stars within 30 pc of the Sun (RL08), plotted against the interstellar radiation
field at 975\AA\ \citep{OpalWeller:1984}.  Vertical dotted lines show
the 975\AA\ radiation field at the solar location.  
The dashed lines show linear fits through the Mg and Fe data points
(see text).  Middle: Fractional ionization for the assumption that all LIC sightlines
have identical true depletions of Fe and Mg, with an additional unseen hydrogen component,
perhaps as H$^+$.  Bottom: the LIC cloud temperature
in each sightline from RL08.  Open circles indicate LIC stars
in third and fourth galactic quadrants, $\ell > 180^\circ$, where
interstellar opacities are small.}
%\end{figure}
\end{wrapfigure}

%% file: frischtable1-100813-astroph.tex
\begin{table}[t!]
\caption{Cloud velocity vectors$^{A}$\label{tab:ismf2} } 
\centering 
\begin{tabular}{l cc cc} 
\\  
\hline  
{Cloud} & {V$_\mathrm{LSR}$} & L$_\mathrm{LSR}$ & {V$_\mathrm{HC}$} &  {L$_\mathrm{HC}$} \\  
{} & {} & {B$_\mathrm{LSR}$} & {} &  {B$_\mathrm{HC}$}  \\  
{} & {(km s$^{-1}$)} & {(deg.)} & {(km s$^{-1}$)} &  {(deg.)}  \\  
\hline  
\\
%%% GCRS1apj_schwadron2d-colorcorrected-myrtlebeach-reformt.pro
%%% Tue Oct  8 20:57:21 2013
   LIC  & $ 15.3 \pm   2.5$  & $141.9 \pm   9.7$  & $ 23.8 \pm   0.9$  & $187.0 \pm   3.4$   \\
        &            & $  6.4 \pm   9.5$  &            & $-13.5 \pm   3.3$   \\
     G  & $ 19.6 \pm   2.2$  & $148.8 \pm   7.5$  & $ 29.6 \pm   1.1$  & $184.5 \pm   1.9$   \\
        &            & $ -9.3 \pm   8.0$  &            & $-20.6 \pm   3.6$   \\
  Blue  & $  7.1 \pm   2.3$  & $ 94.2 \pm  22.0$  & $ 13.9 \pm   0.9$  & $205.5 \pm   4.3$   \\
        &            & $ 17.4 \pm  24.8$  &            & $-21.7 \pm   8.3$   \\
   Aql  & $ 46.6 \pm   2.4$  & $163.4 \pm   4.8$  & $ 58.6 \pm   1.3$  & $187.0 \pm   1.5$   \\
        &            & $-54.9 \pm   2.5$  &            & $-50.8 \pm   1.0$   \\
   Eri  & $ 12.3 \pm   2.2$  & $152.9 \pm  10.8$  & $ 24.1 \pm   1.2$  & $196.7 \pm   2.1$   \\
        &            & $ -0.3 \pm  10.1$  &            & $-17.7 \pm   2.6$   \\
   Aur  & $  9.5 \pm   2.3$  & $185.4 \pm  13.6$  & $ 25.2 \pm   0.8$  & $212.0 \pm   2.4$   \\
        &            & $  1.0 \pm  14.8$  &            & $-16.4 \pm   3.6$   \\
Hyades  & $ 15.4 \pm   2.9$  & $ 87.0 \pm  10.5$  & $ 14.7 \pm   0.8$  & $164.2 \pm   9.4$   \\
        &            & $-10.2 \pm   9.6$  &            & $-42.8 \pm   6.1$   \\
   Mic  & $ 16.3 \pm   2.4$  & $174.2 \pm  10.8$  & $ 28.4 \pm   0.9$  & $203.0 \pm   3.4$   \\
        &            & $ 20.4 \pm   7.9$  &            & $ -3.3 \pm   2.3$   \\
   Oph  & $ 18.0 \pm   1.9$  & $207.8 \pm   9.2$  & $ 32.2 \pm   0.5$  & $217.7 \pm   3.1$   \\
        &            & $ 25.3 \pm   6.6$  &            & $  0.8 \pm   1.8$   \\
   Gem  & $ 22.6 \pm   2.3$  & $191.8 \pm   5.8$  & $ 36.3 \pm   1.1$  & $207.2 \pm   1.6$   \\
        &            & $ 16.7 \pm   5.2$  &            & $ -1.2 \pm   1.3$   \\
   NGP  & $ 26.2 \pm   2.6$  & $166.8 \pm   5.4$  & $ 37.0 \pm   1.4$  & $189.8 \pm   1.7$   \\
        &            & $  8.3 \pm   3.9$  &            & $ -5.4 \pm   1.1$   \\
   Leo  & $ 14.4 \pm   2.6$  & $146.8 \pm  11.8$  & $ 23.5 \pm   1.6$  & $191.3 \pm   2.8$   \\
        &            & $ 14.6 \pm   7.8$  &            & $ -8.9 \pm   1.8$   \\
   Dor  & $ 46.5 \pm   2.0$  & $129.9 \pm   3.4$  & $ 52.9 \pm   0.9$  & $157.3 \pm   1.5$   \\
        &            & $-43.6 \pm   2.3$  &            & $-47.9 \pm   0.6$   \\
   Vel  & $ 30.6 \pm   2.4$  & $176.6 \pm   1.9$  & $ 45.2 \pm   1.8$  & $195.4 \pm   1.1$   \\
        &            & $-13.8 \pm   2.7$  &            & $-19.1 \pm   1.0$   \\
   Cet  & $ 45.9 \pm   3.1$  & $186.6 \pm   2.5$  & $ 60.0 \pm   2.0$  & $197.1 \pm   0.6$   \\
        &            & $ -2.3 \pm   1.8$  &            & $ -8.7 \pm   0.5$   \\
\hline
\vspace*{0.05in}
\end{tabular}  
\flushleft  
\vspace*{-0.2in}  
\footnotesize{\tiny} 
 $^{A}${Cloud heliocentric cloud velocities are from
Redfield \& Linsky (2008).  These velocities are converted to the LSR
using the solar apex motion in Schonrich, Binney and Dehnen (2010) of
U$=11.1 ^{+0.69}_{-0.75}$, V=$12.24^{+0.47}_{-0.47}$, 
W$= 7.25^{+0.37}_{-0.36}$, which gives 
a solar velocity of $18.0 \pm 0.9$ km s$^{-1}$ towards the
direction L=$47.9^\circ \pm 2.9^\circ$, B=$23.8^\circ \pm 2.0^\circ$. The cloud LSR uncertainties include
the uncertainties on the solar apex motion.}
\end{table} 